\documentstyle[eqsecnum,aps]{revtex}
\begin{document}
\draft
\title{
de Broglie-Bohm Interpretation for 
Analytic Solutions of the Wheeler-DeWitt Equation 
in Spherically Symmetric Space-time} 
\author{Masakatsu Kenmoku
\footnote{e-mail address: kenmoku@phys.nara-wu.ac.jp}}
\address{Department of Physics, Nara Women's University, Nara 630-8506, 
Japan}
\author{Hiroto Kubotani
\footnote{e-mail address: kubotani@yukawa.kyoto-u.ac.jp}}
\address{Faculty of Engineering, Kanagawa University, Yokohama 221-8686, 
Japan}
\author{Eiichi Takasugi
\footnote{e-mail address: takasugi@het.phys.sci.osaka-u.ac.jp}}
\address{Department of Physics, Osaka University, Osaka 560-0043, Japan}
\author{Yuki Yamazaki
\footnote{e-mail address: yamazaki@phys.nara-wu.ac.jp}}
\address{Graduate School of Human Culture, Nara Women's
University, Nara 630-8506, Japan}
\date{\today}
\maketitle
\begin{abstract}
We discuss the implications of a wave function for quantum gravity, 
which involves nothing but 3-dimensional geometries as arguments  
and is invariant under general coordinate transformations.
We derive an analytic wave function from the Wheeler-DeWitt equation for 
spherically symmetric space-time with the coordinate system arbitrary.
The de Broglie-Bohm interpretation of quantum mechanics is 
applied to the wave function.
In this interpretation, 
deterministic dynamics can be yielded from a wave function 
in fully quantum regions as well as in semiclassical ones.
By introducing a coordinate system additionally, 
we obtain a cosmological black hole picture 
in compensation for the loss of general covariance.
%as the result of the trade-off of loss of the general covariace.
Our analysis shows that the de Broglie-Bohm interpretation gives quantum 
gravity an appropriate prescription to introduce coordinate systems 
naturally and extract information from 
a wave function as a result of breaking general covariance.
%This shows that the quantum gravity does not reveal implications
%of the wave function until 
%the de Broglie-Bohm prescription breaks its general covariance.
\end{abstract}
%%{\it keywords}:
%%de Broglie-Bohm interpretation, Wheeler-DeWitt equation, 
%%quantum gravity, black hole 
%%%%%%%%%%%%%%%%%%%%%%%%%%%%%%%%%%%%%%%%%%%%%%%%%%%%%%%%%%%%%
\section{Introduction}
%%%%%%%%%%%%%%%%%%%%%%%%%%%%%%%%%%%%%%%%%%%%%%%%%%%%%%%%%%%%%

Quantum theory of gravity is one of the most attractive subjects 
in particle physics and cosmology.
The canonical formalism of gravity formulated by 
Arnowitt, Deser, and Misner (ADM) \cite{ADM62} and by Dirac \cite{Dirac64} 
describes the dynamics of gravity as a totally constraint system.
For quantization, the constraints are used to construct possible quantum 
states.
The equation which corresponds to the Hamiltonian constraint is called 
the Wheeler-DeWitt equation \cite{Wheeler68,DeWitt67}.
A great deal of effort has been made to analyze this equation
\cite{Kuchar93}.
Especially the dynamics of spherically symmetric geometries has been 
studied extensively \cite{Fischler90,Kuchar94,Spherically}.
In this case, although gravitational and electromagnetic 
waves cannot be realized, 
we can treat heuristic geometrical structures of space-time: 
black holes as small scale structures and expanding 
universe as large scale one.

In solving the Wheeler-DeWitt equation, there arise the problem of operator ordering ambiguity  and the regularization problem
\cite{DeWitt67,Schwinger63,TWFJ}. 
It has been argued that although the constraint algebra does close 
in certain operator ordering, it does not necessarily close in 
another ordering.
In our previous paper\cite{Kenmoku99}, 
we considered spherically symmetric geometry and fixed operator ordering 
in the Hamiltonian, momentum, and mass constraints 
so that the algebra among these constraints is closed.
For the consistent ordering, we also found an analytic wave function of 
the quantized spherically symmetric space-time as a simultaneous solution 
of the constraint equations.

To proceed further, 
in this paper, we try to extract quantum properties of the space-time 
from the analytic wave function\cite{Kenmoku99,Yamazaki98}.
For this aim, we use the de Broglie-Bohm (dBB) interpretation 
(quantum potential interpretation or pilot wave approach)
of quantum mechanics\cite{Bohm52,Bell87,Holland93}. 
In this interpretation, a deterministic rigid trajectory 
in the configuration space, 
what we call {\it a de Broglie-Bohm trajectory}, is defined well.
The interpretation merely assigns 
the gradient of the phase of a wave function to 
the momentum of a quantum particle as
\begin{equation}
m\frac{d x}{d t}=p
                \equiv\frac{\partial \Theta}{\partial x},
\label{momentum}
\end{equation}
where the wave function is expressed as 
$\Psi=\mid \Psi \mid \exp{(i\Theta)}$.
A dBB trajectory is obtained by integrating Eq. (\ref{momentum}) with 
respect to $t$. 
There can exist many trajectories specified by the 
configuration on which the quantum system is set at an initial time.
The amplitude $\mid \Psi \mid^2$ is interpreted as  
the probability density in a statistical ensemble of the trajectories. 
For the potential problem of quantum particles, 
the equation to determine the phase of a wave function is derived 
from the Schr\"{o}dinger equation.
It is described as a form of the Hamilton-Jacobi type equation, 
which distinctively includes  
{\it a quantum potential }
\begin{equation}
V_Q=
-  \frac{\hbar^2}{2\mid \Psi \mid}
  \frac{\partial ^2 }{\partial x^2}\mid \Psi \mid.
\end{equation}
A similar equation is derived also for gravity from the Wheeler-DeWitt 
equation.

The dBB interpretation of quantum mechanics is favorable especially 
for quantum theory of gravity because this interpretation 
is able to resolve the conceptual  
problems of quantum gravity: the disappearance of the dynamics 
(the loss of time) \cite{Isham92} and the problem of the observation
\cite{Halliwell87}. 
The dBB interpretation introduces the 
time parameter $t$ through Eq. (\ref{momentum}), 
 which  surmounts the problem of time 
 \cite{Alwis94,Kenmoku96,Kenmoku98}. 
Further, a rigid trajectory can be traced 
without any observation even in fully quantum regions  
 and connects the regions smoothly to semi-classical regions, 
where the quantum potential is merely negligible and the 
trajectory behaves like a classical path.

This paper is organized as follows.
In Sec. 2, we prepare for the main purpose of this paper.
First we perform canonical quantization of 
the Einstein-Maxwell theory with a cosmological 
constant in spherically symmetric space-time. 
Then we derive an analytic solution of the Wheeler-DeWitt equation.
For the space-time with no electromagnetic field, 
we obtained one in our previous paper \cite{Kenmoku99}.
Sec. 3 is  devoted to introducing the dBB interpretation to 
the analytic wave function. 
In Sec. 4, we present the explicit expressions of
the dBB trajectories.
Here we have to give coordinate conditions. 
Summary of this paper is given in Sec. 5.

%%%%%%%%%%%%%%%%%%%%%%%%%%%%%%%%%%%%%%%%%%%%%%%%%%%%%%%%%%
\section{Canonical quantization of the Einstein-Maxwell theory  
in spherically symmetric space-time }
%%%%%%%%%%%%%%%%%%%%%%%%%%%%%%%%%%%%%%%%%%%%%%%%%%%%%%%%%%%

In this section we consider canonical quantization of 
the 4-dimensional Einstein-Maxwell theory with a cosmological term. 
Space-time is assumed to be spherically symmetric.
We quantize the space-time following our previous work \cite{Kenmoku99}, 
where we fixed operator ordering in the Hamiltonian, momentum, 
and mass constraints and showed that they form a closed algebra. 
We also derive an analytic wave function 
which satisfies these constraints.
Here we extend our previous model \cite{Kenmoku99} to include 
the electromagnetic field.
The inclusion of the electromagnetic field will be profitable  
to examine extreme black holes \cite{extreme} 
and cosmological black holes \cite{Kastor93}. 
We use natural geometrical units $c=\hbar=G=1$ and adopt 
the conventions in the Kucha$\check{\rm r}$'s work \cite{Kuchar94} 
and ours\cite{Kenmoku99}.

\subsection{Canonical quantization}

We start to consider a general spherically symmetric metric 
in the ADM decomposition,
\begin{equation}
    ds^2 = -N^2 dt^2 +\Lambda^2(dr + N^{r} dt)^2 + R^2 d\Omega^2,
    \label{ds^2}
\end{equation}
where $d\Omega$ is a line element on the unit sphere, 
and the metric components $N$, $N^{r}$, $\Lambda$, and $R$ are functions of 
the time coordinate $t$ and the radial one $r$.
Hereafter we refer to $N$ and $N^r$ as the lapse and shift functions, 
respectively, 
which express the degrees of freedom in the choice of coordinate systems.
The action of the Einstein-Maxwell theory has a form of
\begin{equation}
    I = \frac{1}{16\pi} \int d^4x \sqrt{-^{(4)}g}
        (^{(4)}R - 2\lambda -F_{\mu \nu}F^{\mu \nu}),
\label{I}
\end{equation}
where $\lambda$ denotes a cosmological term,
and $^{(4)}R$ and $\ ^{(4)}g $ are the scalar curvature and  
the determinant of the metric tensor, respectively. 
Here the electromagnetic field strength is denoted by $F_{\mu \nu}$.
However, spherical symmetry reduces the number of 
non-vanishing components of $F_{\mu \nu}$.
The nontrivial part is described as 
$F_{01} =-F_{10}= \dot{A_1} - {A_0}'$,
where $A_0$ and $A_1$ are the components of the electromagnetic field. 
Hereafter a dot and a prime 
denote the derivatives with respect to $t$ and $r$, respectively.
Substituting the metric (\ref{ds^2}) into Eq. (\ref{I}), 
we transform the action into a form of the ADM decomposition
\begin{eqnarray}
I&=&\int dt \int dr [
-N^{-1} \Bigl( R(-\dot \Lambda + (\Lambda {N^{r}} )') (-\dot R + R' N^{r})
     + \frac{1}{2} \Lambda (-\dot R + R' N^{r})^2 \Bigr)
       \nonumber \\ 
& &+N\Bigl(-\Lambda^{-1}RR''-\frac{1}{2}\Lambda^{-1}R'^2
     +\Lambda^{-2}\Lambda ' RR'
       +\frac{1}{2} \Lambda(1-\lambda R^2)\Bigr)
       \nonumber \\
    & &+ \frac{1}{2} N^{-1} \Lambda ^{-1} R^2  (\dot{A_1}-A_0')^2                          ] \  \label{I-ADM} \\
    &=& \int dt \int dr [
    P_\Lambda \dot \Lambda +P_R \dot R +P_A \dot A_1
    -N \Bigl( \frac{1}{2}\Lambda R^{-2}P_\Lambda^2-R^{-1}P_R P_\Lambda
    +\frac{1}{2}R'^2\Lambda^{-1} \nonumber \\
    & &+R R''\Lambda ^{-1} -R R'\Lambda^{-2}\Lambda '
    -\frac{\Lambda}{2}(1-R^{-2}P_A^2 -\lambda R^2)\Bigr)
    \nonumber \\
    & &-N^r \Bigl( R' P_R -\Lambda (P_{\Lambda})' \Bigr)-A_0 \hat P_A'],
\label{I-ADM2}
\end{eqnarray} 
where $P_\Lambda$, $P_R$, and $P_A$ are the canonical momenta 
conjugated to $\Lambda, R$, and $A_1$, respectively.

We quantize the dynamical variables $\Lambda, R$, and $A_1$. 
In the Schr\"{o}dinger picture, their canonical momenta are represented 
by functional differential operators: 
\begin{eqnarray}
\hat{P}_{\Lambda} (r) &=& -i \frac{\delta}{\delta \Lambda (r)},\nonumber \\
\hat{P}_R    (r) &=& -i \frac{\delta}{\delta R   (r)} \ ,
    \label{S-representation} \\
    \hat{P}_A    (r) &=& -i \frac{\delta}{\delta A_1 (r)} \ . \nonumber
\end{eqnarray}
Here and in the rest of this section, 
we use the notation {\lq\lq}hat" for differential operators 
and do not express the argument $t$ explicitly, 
because we always treat products of simultaneous operators only. 
%operators to which the same time is assigned.

\subsection{Constraint equations}

The action (\ref{I-ADM2}) includes the non-dynamical variables 
$N, N^r$, and $A_1$.
Variation with respect to them induces the classical constraint equations: 
the Hamiltonian and momentum constrains and the Gaussian law.
The constraints are adapted to restrict a wave function of  
quantum gravity, $\Psi$ in a form of the equations 
\begin{eqnarray}
\hat{H}\Psi      &=& 0, \nonumber\\      
\hat{H}_{r}\Psi &=& 0, \label{constraint} \\
\hat{H}_A\Psi  &=& 0, \nonumber
\end{eqnarray}
where the concrete expressions for the operators 
$\hat{H}, \hat{H}_{r}$, and $\hat{H}_A$ are given later.
The first equation is called the Wheeler-DeWitt equation, and 
the third one corresponds to the conservation of electric charge.
For spherically symmetric space-time with no matter, 
we can introduce the mass of a black hole, $M$ as a dynamical variable
\cite{Fischler90,Kuchar94,Nambu88}, which is a constant of motion.
In quantum theory, we can construct mass eigenstates through
\begin{equation}
    \hat{M} \Psi     = m\Psi, 
\label{mass equation}
\end{equation}
where $\hat{M}$ and $m$ are 
the quantized mass operator and a mass eigenvalue, respectively.

%\subsection{Operator ordering in constraint equations}

In quantum theory of gravity, 
it is troublesome to fix operator ordering.
First we introduce the Hamiltonian, momentum, electric charge,  
and mass operators with ordering factors as 
\begin{eqnarray}
\hat H&=&\frac{1}{2}\Lambda R^{-2}\hat P_\Lambda^{(C)}\hat P_\Lambda
-R^{-1}\hat P_R\Lambda \hat P_\Lambda^{(B)}\Lambda^{-1}  
+\frac{1}{2}R'^2\Lambda^{-1} 
\nonumber\\ 
&&
+ R R''\Lambda ^{-1} 
-R R'\Lambda^{-2}\Lambda '
-\frac{\Lambda}{2}\left (1-\hat P_A^2 R^{-2} -\lambda R^2 \right ),
\label{Hamiltonian operator}\\
\hat{H}_r  &=& R' \hat{P}_R -\Lambda (\hat{P}_{\Lambda})',
\label{Momentum operator}\\
\hat H_A&=& -(\hat P_A)',
\label{Charge operator}\\
\hat M-m&=&\frac{1}{2}R^{-1} \hat P_\Lambda^{(A)} \hat P_\Lambda 
        -\frac{1}{2} R (\chi -\hat F),
\label{Mass operator} 
\end{eqnarray}
where  
\begin{eqnarray}
\chi &\equiv& R'^2 ~ \Lambda^{-2}  \;,\label{chi}\\
\hat F&\equiv&1-2mR^{-1} + \hat P_A^2R^{-2}
-\frac{\lambda}{3}R^2.
\label{F}
\end{eqnarray}
In Eqs. (\ref{Hamiltonian operator}) and (\ref{Mass operator}), 
the momentum $\hat P_\Lambda$ is accompanied with an ordering 
function $A$ as
\begin{eqnarray}
\hat P_\Lambda^{(A)}
         &\equiv& A\hat P_\Lambda A^{-1}, \nonumber\\
\hat P_\Lambda^{(B)}
         &\equiv&\frac{1}{2}\left(\hat P_\Lambda+\hat P_\Lambda^{(A)}\right)
          =A^{1/2}\hat P_\Lambda A^{-1/2}, \label{orderingP} \\
\hat P_\Lambda^{(C)}
         &\equiv&\hat P_\Lambda^{(A)}-iR R'^{-1}
            \left( A^{-1}\frac{\delta A}{\delta \Lambda} \right)'
         =C \hat P_\Lambda C^{-1}, \nonumber
\end{eqnarray}
where 
\begin{equation}
C=A \exp \left[-\int^r dr RR'^{-1}\int^\Lambda d\Lambda\left (A^{-1}
\frac{\delta A}{\delta \Lambda}\right)'\right ]\;.
\end{equation}
In this ordering, $\hat M$ has a favorable relation 
\begin{equation}
\hat M '=-\Lambda^{-1}R'\hat H-   R^{-1}\hat P_\Lambda^{(B)}
\Lambda^{-1}\hat H_r -R^{-1} \hat P_A \hat H_A\;,
\label{dotM}
\end{equation} 
which guarantees that $\hat M$ is spatially conserved as well as 
in classical theory.
It shows that we have provided the quantized constraints 
(\ref{Hamiltonian operator}) - (\ref{Mass operator}) with 
consistent ordering.

Next we fix the ordering function $A$ so that $\hat{H}, \hat{H}_{r}$, and 
$\hat M$ forms a closed algebra.
According to \cite{Kenmoku99}, 
the commutators between $\hat{H}$ and $\hat{H}$ or $\hat{H}_r$ are evaluated 
from the commutators between $\hat M$ and $\hat M$ or $\hat{H}_r$, 
and $A$ is shown to take a form
\begin{equation}
A=A_Z(Z)\bar A(R,\chi)\;,
\label{orderingA}
\end{equation}
where $A_Z$ and $\bar A$ are arbitrary functions, and 
the argument $Z$ is defined as
\begin{equation}
    Z\equiv\int dr  \Lambda f(R,\chi)=\int dr 
    \int^{\Lambda}d\Lambda \bar f(R,\chi)
    \label{Z}
\end{equation}
by an arbitrary function, $f$ or $\bar f$.
Here $f$ and $\bar f$ have a relation 
\begin{equation}
f(R, \chi)=- \frac{\chi^{1/2}}{2} \int^{\chi} dx
x^{-3/2} \bar f(R,x) \;. \nonumber
\end{equation}
In the proof of the closure of the constraint algebra, 
we make the most of a special property that 
\begin{equation}
 [Z, H_r(r)]=i\left( R'(r)\frac{\delta Z}{\delta R(r)} 
 - \Lambda (r)(\frac{\delta Z}{\delta \Lambda (r)})'\right )=0.
\label{Zbracket}
\end{equation}

Tsamis and Woodard \cite{TWFJ} pointed out that the closure of the 
constraints in quantum gravity  
is ill-defined in the sense that formal evaluation of the commutator 
yields the multiple product of delta functions, and that
the theory must be regulated.
Especially, the formal adoption of calculation rules for  
the delta function to the the multiple product of delta functions 
with the same argument leads even to contradictory results.
In our evaluation of the commutator\cite{Kenmoku99}, on the other hand, 
we only have to utilize the commutators of $\hat M$ and 
thus need not treat the multiple product of delta functions with 
the same argument.
Only multiple products of delta functions in our calculation 
 comprise $\delta(0)$ that comes from the commutation of 
coincident operators.
Further, all the noncancelable ones are multiplied by the constraints in 
the canonical gravity.
Thus, we hopefully proceed further through 
the formal manipulation of the unregulated theory.

\subsection{Solutions of constraint equations}

The wave function $\Psi$ is a functional of 
the electromagnetic field $A_1$ and the geometrical fields  
$\Lambda$ and $R$. 
The wave function is assumed to be in a separable form 
\begin{equation}
   \Psi = \Psi_{EM}[A_1] \Psi_G[\Lambda,R] \ .
   \label{Psi}
\end{equation}
Each of $\Psi_{EM}[A_1]$ and $\Psi_G[\Lambda,R]$ is constructed as follows.
The quantized Gaussian law $\hat{H_A}\Psi_{EM}[A_1]=0$
is trivially satisfied by the equation
\begin{equation}
\hat{P}_A\Psi_{EM} = Q \Psi_{EM},
\end{equation}
whose solution is
\begin{equation}
\Psi_{EM}[A_1] = \exp (i \int dr \ Q A_1(r)).
\label{PsiA} 
\end{equation}
Here $Q$ is an eigenvalue of the conserved charge.
Considering Eq. (\ref{Zbracket}), a solution of the momentum constraint 
$\hat H_r \Psi_G[\Lambda,R]=0$ is 
\begin{equation}
    \Psi_G = \Psi_G(Z),
    \label{PsiG}
\end{equation}
where $\Psi_G(Z)$ on the right hand side is an arbitrary function of $Z$.

From Eq. (\ref{dotM}), 
we see that if  $\Psi_G(Z)$ satisfies the mass constraint 
(\ref{mass equation}),
it simultaneously satisfies the Wheeler-DeWitt equation.
Thus we only have to consider the mass constraint instead of 
the Wheeler-DeWitt equation.
As to the ordering function $\bar A$ in Eq. (\ref{orderingA}), 
we assume that 
\begin{equation}
\frac{\delta Z}{\delta \Lambda}(\equiv\bar{f})=\bar A
{\rm ~~~and~~~}\bar A^2=R^2(\chi-F(R)),
\label{orderingAbar}
\end{equation}
where $F\equiv\hat F(\hat P_A\to Q)$.
This assumption reduces the differential functional 
equation (\ref{mass equation}) with Eq. (\ref{Mass operator}) 
to an ordinary differential equation on $Z$, 
\begin{equation}
 \frac{d^2\Psi_G}{d Z^2}
-{1 \over A_Z}\frac{d A_Z}{d Z} \frac{d \Psi_G}{d Z}
+\Psi_G =0\;.
\label{ordinarydif}
\end{equation}
The variable $Z$ in Eq. (\ref{Z}) is also determined as
\begin{eqnarray}
Z&=&\int d r \int ^{\Lambda} d\Lambda \ R \sqrt{\chi - F(R)} \nonumber\\
 &=& \int dr R\Lambda \left( \sqrt{\chi - F(R)} + 
         \frac{\sqrt{\chi}}{2}  \ln \left | 
          \frac{\sqrt{\chi - F(R)}-\sqrt{\chi} 
          }{\sqrt{\chi - F(R)}+\sqrt{\chi}} \right | \right)  \ . 
\label{Zint}
\end{eqnarray}

Finally we choose the rest of the ordering factors, $A_Z$ such that 
the solution of Eq. (\ref{ordinarydif}) becomes a special function.  

\vskip 1mm
\noindent
(i) Bessel type solutions
\vskip 1mm
\noindent
As the simplest case, we take 
\begin{equation}
A_Z=Z^{2\nu-1}.
\end{equation}
Then the solution of Eq. (\ref{ordinarydif}) is given by 
the Hankel (or Bessel) functions as
\begin{equation}
    \Psi_G^{(\nu)}(Z) = Z^{\nu} \ 
                (a_1 \ H_{\nu}^{(1)}(Z)+a_2 \ H_{\nu}^{(2)}(Z))
        \ ,
        \label{wf1}
\end{equation}
where  $a_1$ and $a_2$ are integration constants.

\vskip 1mm
\noindent
(ii) Hypergeometric type solutions

\vskip 1mm
\noindent
If we choose 
\begin{equation}
A_Z=Z^{\sigma}(Z-1)^{\delta},
\end{equation} 
the solution of Eq. (\ref{ordinarydif}) is given by the hypergeometric 
functions as
\begin{eqnarray}
\Psi_G^{(\sigma,\delta)}(Z)&=&Z^{\sigma +1}(Z-1)^{\delta +1}
\Bigl\{ b_1 F(\alpha,\beta,\gamma;Z) \Bigr.  \nonumber\\
& &\quad\Bigl.  
+b_2 Z^{1-\gamma}F(\alpha-\gamma+1, \beta-\gamma+1,2-\gamma;Z)
 \Bigr\},
\label{wf2}
\end{eqnarray}
where $b_1$ and $b_2$ are integration constants, and $\alpha$, $\beta$, 
and $\gamma$ are constants under the constraints 
\begin{equation}
\alpha\beta=\sigma+\delta+2\;,\;\;\alpha+\beta=\sigma+\delta+3\;,
\;\;\gamma=\sigma+2\;.
\end{equation}

\vskip 1mm
Each function of Eqs. (\ref{wf1}) and (\ref{wf2}) shows an eigenstate of 
the mass $m$.
Eigenstates with different mass eigenvalues can be superposed.
We also mention generality of the wave functions 
(\ref{wf1}) and (\ref{wf2}).
For the Hamilton-Jacobi equation, a complete solution contains 
the same number of arbitrary constants as of the dynamical variables.
In our model, there are two dynamical variables $R$ and $\Lambda$ 
at each point of $r$.
However, the Wheeler-DeWitt equation works
as if it is a kind of Hamilton-Jacobi equation with zero energy. 
Therefore, the zero energy condition is forced at each point of $r$.
Further, we set the mass constraint equation (\ref{mass equation}) 
at each point.
As a result, our solutions contain only the universal mass eigenvalue as 
an arbitrary constant 
unlike a complete solution of the Hamilton-Jacobi equation.

%%%%%%%%%%%%%%%%%%%%%%%%%%%%%%%%%%%%%%%%%%%%%%%%%%%%%%%%%%%%%%%%%
%
\section{de Broglie-Bohm interpretation}
%
%%%%%%%%%%%%%%%%%%%%%%%%%%%%%%%%%%%%%%%%%%%%%%%%%%%%%%%%%%%%%%%%%
In the previous section, we obtained the wave function for 
spherically symmetric space-time.
It should be noted that, 
although some assumptions are imposed in order to get the analytic formula, 
any coordinate condition on $N$ and $N^r$ has not been undertaken.
For classical relativity, 
we need to fix a coordinate system to get an explicit expression 
of a space-time geometry. 
By using the dBB interpretation of quantum mechanics, 
we try to extract physical meanings 
from the wave function of quantum gravity.
For ordinary quantum systems, as briefly seen in Sec. 1,  
the dBB interpretation gives us deterministic rigid trajectories  
with no ambiguity instead of wave functions.
In this section, we shows how the dBB interpretation 
induces a rigid space-time picture in quantum theory of gravity.

We express a wave function of quantum gravity 
in a polar coordinate as     
\begin{equation}
     \Psi(Z) =\mid \Psi(Z) \mid \exp{(i\Theta(Z))},
\label{polar}
\end{equation}
where the phase $\Theta(Z)$ depends only on $Z$ as well as the amplitude 
$|\Psi(Z)|$.
In a similar way to Eq. (\ref{momentum}), the derivatives of the phase $\Theta$ 
are identified with the canonical momenta $P_\Lambda$, $P_R$, 
and $P_A$ which are conjugate respectively to $\Lambda$, $R$, and $A_1$. 
Estimating the derivatives as 
\begin{eqnarray}
    \frac{\delta \Theta}{\delta \Lambda}
         &=& \frac{\delta Z}{\delta \Lambda}\frac{d \Theta}{dZ}
         = \bar{f} \ \frac{d \Theta}{dZ}\ , \nonumber  \\
    \frac{\delta \Theta}{\delta R}
         &=& \frac{\delta Z}{\delta R}\frac{d \Theta}{dZ}
         = \frac{\Lambda}{R'}(\frac{\delta Z}{\delta \Lambda})'
           \ \frac{d \Theta}{dZ}
         = \frac{\Lambda}{R'}\bar{f}' 
           \ \frac{d \Theta}{dZ}\ ,  \label{Derivatives} \\
      \frac{\delta \Theta}{\delta A_1}
         &=& Q\ ,\nonumber
\end{eqnarray} 
we obtain the equations to determine the dBB trajectory of the 
space-time geometry and the electromagnetic field as 
\begin{eqnarray}
     \dot{R}-R'N^r 
           &=&-\frac{N}{R}\bar f \frac{d \Theta}{d Z} \ , 
           \label{dBB R}
           \\  
     R(\dot{\Lambda}-(\Lambda N^r)')
           +\Lambda(\dot{R}-R'N^r) &=&-\frac{N \Lambda}{R'}
           \bar{f}' \frac{d \Theta}{d Z} \ ,    
           \label{dBB L} \\
     \dot{A}_1 - A_0' 
           &=& Q N \Lambda R^{-2} \ ,
           \label{dBB A}
\end{eqnarray} 
where $\bar f= R\sqrt{\chi-F(R)}$ from the assumption (\ref{orderingAbar}) 
and  $\chi\equiv R'^2\Lambda^{-2}$ by the definition (\ref{chi}).
In the estimation (\ref{Derivatives}), we have  used Eq. (\ref{Zbracket}).
Equations (\ref{dBB R})-(\ref{dBB A}) form simultaneous differential 
equations with respect to the time and radial coordinates.
By comparing a solution of the equations with the classical one 
which is derived from the classical equation of motion, 
we can know quantum gravity effects on the space-time geometry 
qualitatively.
When the difference between them is negligible, 
we only have to assert that the quantum space-time is reduced to 
the classical one spontaneously, 
or the classical state is realized without any observer.

Before integrating Eqs. (\ref{dBB R})-(\ref{dBB A}), 
we observe them for a while.
At sight we find that the equation for the electromagnetic 
part, (\ref{dBB A}) is the same as the classical one. 
By taking the ratio of Eq. (\ref{dBB L}) 
to Eq. (\ref{dBB R}), we obtain 
\begin{equation}
    \frac{\dot{\Lambda}}{\Lambda}+\frac{\dot{R}}{R}
   -\frac{\dot{R}}{R'}{\frac{\bar{f}}{\bar{f}}}'
   = {N^r}'+N^r (\ln \frac{\Lambda R}{\bar{f}})' \ .
\label{rational2}
\end{equation}
%%%%%%%%%%%%%%%%%%%%%%%%%%%%%%%%%%%%%%%%%%%%%%%%%%%%%
%This relation is independent on the explicit functional 
%form of the phase factor and thus the quantum potential 
%that usually enters though the phase.
%%%%%%%%%%%%%%%%%%%%%%%%%%%%%%%%%%%%%%%%%%%%%%%%%%%%%
We note that Eq. (\ref{rational2}) doesn't include 
the phase $\Theta$ and the solution of Eq. (\ref{rational2}) will not 
depend on an explicit form of the wave function.
For the gravitational part, thus the correlation between 
the dynamical variables $\Lambda$ and $R$ corresponds to 
the classical path in the configuration space.
By contrast to Eq. (\ref{rational2}), Eq. (\ref{dBB R}) includes 
a functional form of $\Theta$ and the time slicing function $N$.
Combined with Eq. (\ref{rational2}), Eq. (\ref{dBB R}) is used 
to determine finally the dBB trajectory of the space-time geometry  
which is parametrized by the time $t$.

In order to proceed to get dBB trajectories concretely, 
we specify a wave function for our analysis.
We adopt the Bessel type solution (\ref{wf1}) with $a_1=0$, 
\begin{equation}
\Psi_G^{(\nu)}(Z)=a_2 Z^{\nu}H_{\nu}^{(2)}(Z),
\label{wf3}
\end{equation}
since its asymptotic behavior is simple and thus it is hopeful to derive 
physical meanings of the wave function.
For the wave function (\ref{wf3}), 
we can estimate $d\Theta/dZ$ analytically by using the relations 
$ H^{(1)}_{\nu}(Z)^*=H^{(2)}_{\nu}(Z)$ on the real $Z$
and 
$H^{(2)}_{\nu}(dH^{(1)}_{\nu}/dZ)
  - H^{(1)}_{\nu}(dH^{(2)}_{\nu}/dZ)= 4i/(\pi Z)$.
The result is  
\begin{equation}
n(Z)^{-1}\equiv -\frac{d \Theta}{dZ}
        =\frac{2}{\pi Z \mid H^{(2)}_{\nu}(Z)\mid ^2 }.
\label{asymDefTheta}
\end{equation}
Thus $n(Z)$ is asymptotically 
\begin{equation}
n(Z)\; \to \; 1 
    \qquad {\rm for}\quad Z \to \infty   \;.
\end{equation}
Although $\Psi_G^{(\nu)}(Z)$ is an analytic function on the complex $Z$,
the phase $\Theta(Z)$ is not.
Therefore, we confine $n(Z)$ on the real $Z$ and refrain from 
extending $n(Z)$ analytically to the complex plane.
We note that the wave function (\ref{wf3}) satisfies 
the Vilenkin's boundary condition \cite{Vilenkin88}, 
since the time derivative of $R$ has a positive sign as seen 
in Eq. (\ref{dBB R}).
It may be worthwhile to mention other choices of the wave function.
If $b_1=b_2$ or $b_1=-b_2$, the wave function remains real 
or purely imaginary, and the dBB interpretation gives no dynamics.
For the case $b_2=0$, 
the sign of the phase in the wave function is opposite to 
that in the wave function (\ref{wf3}) and thus 
the direction of the time parameter $t$ is reversed.

%%%%%%%%%%%%%%%%%%%%%%%%%%%%%%%%%%%%%%%%%%%%%%%%
%The dBB equations contains $N$, $N^r$, $R$, $R'$ and 
%$\Lambda$. Therefore, we have to specify $N^r$ (the 
%gauge condition), and synchronize $R$ to $t$ and $r$,  
%{\i.e.}, the condition for $R'$ and $\dot R$ (the 
%coordinate condition). We call these conditions 
%the gauge fixing condition. Then, the lapse function 
%$N$ is determined through the dBB equations. 
%%%%%%%%%%%%%%%%%%%%%%%%%%%%%%%%%%%%%%%%%%%%%%%%

%%%%%%%%%%%%%%%%%%%%%%%%%%%%%%%%%%%%%%%%%%%%%%%%%%%%%%%%%%%%%%%%%%%%
%
\section{de Broglie-Bohm trajectory}
%
%%%%%%%%%%%%%%%%%%%%%%%%%%%%%%%%%%%%%%%%%%%%%%%%%%%%%%%%%%%%%%%%%%%%
In this section, we solve the differential equations 
(\ref{dBB R}) and (\ref{dBB L}) and obtain the dBB trajectories 
hidden in the wave function (\ref{wf3}).
For the equations to be simplified and integrated analytically, 
we assume that $\chi$ defined in Eq. (\ref{chi}) depends on $r$ 
only through $R$:
\begin{equation}
\chi\equiv R'^2/\Lambda^2=\bar{\chi}(R),
\label{barchi}
\end{equation}
where $\bar{\chi}(R)$ is a function of $R$.
The assumption requires that $\bar{f}$ defined in Eq. (\ref{orderingAbar})  
is also a function only of $R$.
If we take a coordinate condition $N^r=0$,
Eq. (\ref{rational2}) is reduced to a simple form 
\begin{equation}
\frac{\dot{\Lambda}}{\Lambda}+\frac{\dot{R}}{R}
   - \frac{\dot{\bar{f}}}{\bar{f}} = 0. 
\label{noNr}
\end{equation}
By integrating Eq. (\ref{noNr}), we obtain a relation 
\begin{equation}
\bar{f}=c_0(r) R\Lambda,
\label{fbar}
\end{equation}
where $c_0$ is an arbitrary function of $r$ as an integration constant 
with respect to $t$. 
From the definition (\ref{orderingAbar}), 
we can determine $\Lambda$ through $R$:
\begin{equation}
\Lambda = \sqrt{\bar{\chi}(R)-F(R)}/c_0(r)\ . 
\label{solution Lambda}
\end{equation}
From Eq. (\ref{barchi}), 
on the other hand, $R'$ is expressed as
\begin{equation}
R'= \sqrt{\bar{\chi}(R)}\Lambda
  =\sqrt{\bar{\chi}(R)(\bar{\chi}(R)-F(R))}/c_0(r)\ .
\label{dashR}
\end{equation}
If the right hand side of Eq. (\ref{dashR}) is nonzero, 
we obtain a solution
\begin{equation}
G(R)\equiv\int {dR \over \sqrt{\bar{\chi}(R)(\bar{\chi}(R)-F(R))}}
=\int {dr \over c_0(r)}+\phi(t),
\label{solution R}
\end{equation}
which determines implicitly the dependence of $R$ on $r$.
Here $\phi$ is an arbitrary function of $t$ as an 
integration constant with respect to $r$. 
The remaining equation of motion (\ref{dBB R}) is used to determine 
the time dependence of $\phi(t)$:
\begin{equation}
\dot \phi
={dG(R) \over dR}\dot R={N \over n(Z)}{1 \over \sqrt{\bar \chi (R)}}.
\label{dot_phi}
\end{equation}
It is noted that, if only we take $N^r=0$ as a coordinate condition,
the ansatz (\ref{barchi}) is compatible with the canonical evolution 
and, resultantly, corresponds to a condition to pick up a special solution.

In the following, we evaluate dBB trajectories by giving explicit
functional forms of $\bar \chi(R)$.
The notations $ds_{dBB}$ and $Z_{dBB}$ are used for 
the line element of the rigid geometry described by the dBB picture  
and $Z$ which is evaluated on the rigid geometry from Eq. (\ref{Zint}), 
respectively.

%%%%%%%%%%%%%%%%%%%%%%%%%%  case A  %%%%%%%%%%%%%%%%%%%%%%%%%%%%
\vspace{3mm}
\noindent
Case A: \  $\bar{\chi}(R)\equiv F(R)\ge 0$ 
\vspace{3mm}

This is a special case where $\bar f=0$ 
and, therefore, Eqs. (\ref{solution Lambda}) and (\ref{solution R}) 
do not apply. 
Directly from  Eqs. (\ref{dBB R}) and (\ref{dBB L}) with $N^r=0$, 
we find 
\begin{equation}
\dot{\Lambda}=\dot{R}=0, 
\end{equation}
which means that there is no dynamical evolution.
This conclusion is confirmed by 
observing that the integrand of $Z$ in Eq. (\ref{Z}) or Eq. (\ref{Zint}) 
vanishes and thus the wave function becomes constant. 
To keep the ansatz $\bar{\chi}(R)\equiv F(R)$, or $R'^2=\Lambda^2 F(R)$ 
during the time evolution we require a condition
\begin{equation}
{\partial \over \partial t}R'=0.
\label{consistent}
\end{equation}
If we take $R'=R$, that is, $R=\exp r$, which satisfies the consistency
condition (\ref{consistent}), then we obtain 
$\Lambda^2=R^2 F(R)^{-1}$ and 
\begin{equation}
ds^2_{dBB}=-N^2 dt^2 +F(R)^{-1}dR ^2+R^2 d\Omega^2.
\end{equation}
Here the lapse function $N$ remains an arbitrary function.
The Reissner-Norstr\"{o}m-de Sitter metric corresponds to 
the case of the choice $N=\sqrt{F(R)}$.

%%%%%%%%%%%%%%%%%%%%%%%%  case B %%%%%%%%%%%%%%%%%%%%%%%%%%
\vspace{3mm}
\noindent
Case B: \ $\bar{\chi}(R)\equiv 0$ 
\vspace{3mm}

First we take the integration constant in Eq. (\ref{fbar}) as $c_0=1$.
From Eq. (\ref{solution Lambda}), 
when $F(R) \le 0$, $\Lambda$ is given by 
\begin{equation}
\Lambda^2=-F(R).
\label{Lambda_B} 
\end{equation}
From Eq. (\ref{dashR}), we have $R'=0$.
Therefore, 
\begin{equation}
R=\phi(t),
\label{R_B}
\end{equation}
where $\phi(t)$ is an arbitrary function of $t$ as an integration constant 
with the respect to $r$.
Equations (\ref{Lambda_B}) and (\ref{R_B}) show an initial configuration 
to pick up a dBB trajectory.

If the lapse function $N$ is taken as
\begin{equation}
N=\frac{1}{\sqrt{-F(R)}} n(Z),
\label{N_B}
\end{equation}
then Eq. (\ref{dBB R}) is reduced to $\dot{\phi}=1$, and 
the dBB trajectory is 
\begin{equation}
R(r,t)=t.
\end{equation}
As a result, the line element $ds_{dBB}$ is 
\begin{equation}
ds^2_{dBB}=\frac{1}{ F(t)} n(Z_{dBB})^2 dt^2 - F(t) dr ^2 + t^2 d\Omega^2,
\label{ds_B}
\end{equation}
where 
\begin{equation}
Z_{dBB}=-\int dr R F(R)= -tF(t)R_0,     
\label{Z_B}
\end{equation} 
and $R_0$ is the world size. 
For large $t$, the metric (\ref{ds_B}) approaches to the inside geometry of 
the Reissner-Nordstr\"{o}m-de Sitter black hole,
which is discussed in Ref. \cite{Yamazaki98},
and, when  $Q=\lambda=0$, to that of 
the Schwarzschild black hole, which is 
discussed in Refs. \cite{Kenmoku98,Nakamura93}.
These classical solutions are generalizations of the 
Kantowski-Sachs metric \cite{Kantowski66}.

%%%%%%%%%%%%%%%%%%%%%%%%%%%% case C  %%%%%%%%%%%%%%%%%%%%%%%%%%%%%%
\vspace{3mm}
\noindent
Case C: \  $\bar{\chi}(R)\equiv 1-2m/R + Q^2/R^2 \ge 0$ 
\vspace{3mm}

The ansatz reduces Eq. (\ref{Zint}) into 
\begin{equation}
Z=  \int dr R^2\left(\sqrt\frac{\lambda}{3}R+
          \frac{\sqrt{\bar \chi(R)}}{2}  \ln \left | 
           \frac{\sqrt{{\lambda}/{3}}\ R-
          \sqrt{\bar \chi(R)}}{\sqrt{{\lambda}/{3}}\ R+
          \sqrt{\bar \chi(R)}
          } \right |\right) \;.
\label{Z_C}
\end{equation}
If we set $c_0(r)=\sqrt{\lambda/3}$ in Eq. (\ref{fbar}), 
then $\Lambda$ is given as 
\begin{equation}
\Lambda= R
\label{Lambda_C}
\end{equation}
from Eq. (\ref{solution Lambda}), and 
Eq. (\ref{dashR}) becomes 
\begin{equation}
R'=R\sqrt{1-{2m \over R}+{Q^2 \over R^2}}.
\label{dashR_C}
\end{equation}
Integrating Eq. (\ref{dashR_C}), we obtain
\begin{equation}
R=x(1+\frac{m}{x}+\frac{m^2-Q^2}{4x^2}),
\label{R_C}
\end{equation}
where $x\equiv\exp{(r+\sqrt{{\lambda \over 3}}\phi(t))}$, 
and $\phi(t)$ is an arbitrary function 
of $t$ as an integration constant with respect to $r$.
Equations (\ref{Lambda_C}) and (\ref{R_C}) describe an initial 
configuration.

Next we choose a time coordinate condition 
\begin{equation}
N=\sqrt{1-{2m \over R}+{Q^2 \over R^2}}~n(Z)
=\Bigl(1-{m^2 -Q^2 \over x^2}\Bigr)
 \Bigl(1+{2m \over x}+{m^2-Q^2 \over x^2}\Bigr)^{-1}~n(Z).
\end{equation}
Then Eq. (\ref{dot_phi}) is reduced to $\dot{\phi}=1$.
Therefore, the parameter $x$ is determined to be
\begin{equation}
x=a_0\exp{(\sqrt{\lambda \over 3} t+r)}\equiv a(t)\rho (r),
\end{equation}
where $a_0$ is a constant, and $a(t)$ corresponds to the scale factor of 
an expanding de Sitter universe with a cosmological term $\lambda$.
As a result, the line element $ds_{dBB}$ is 
\begin{eqnarray}
ds_{dBB}^2 
&=&-\Bigl(1-\frac{2m}{R}+\frac{Q^2}{R^2}\Bigr)n(Z_{dBB})^2dt^2
     +R^2(dr^2+d\Omega^2)\nonumber\\
&=&-\left (\frac{1-\frac{m^2-Q^2}{4x^2}}
       {1+\frac{m}{x}+\frac{m^2-Q^2}{4x^2}}\right)^2 n(Z_{dBB})^2dt^2
     +a(t)^2\ \left(1+\frac{m}{x}+\frac{m^2-Q^2}{4x^2}\right)^2
       (d\rho^2+\rho^2 d\Omega^2).
        \nonumber\\
\label{ds_C} 
\end{eqnarray}

Here we study some limiting cases.
First we consider the special case $ m=Q=0 $. 
Equation (\ref{ds_C}) is reduced to  
\begin{equation}
ds_{dBB}^2=-n(Z_{dBB})^2 dt^2+a(t)^2(d\rho^2+\rho^2d\Omega^2), 
\label{ds_C1}
\end{equation} 
whose asymptotic form behaves like a classical de Sitter universe.
Horiguchi \cite{Horiguchi94} showed that 
quantization of the homogeneous universe with a cosmological term gives 
a Bessel type wave function of order $\nu = 1/3$ (the Airy function)
with the argument
\begin{equation}
    Z= \int d\rho \rho^2 a(t)^3 = a(t)^3 V_0 \ ,
\label{Z_Ca}
\end{equation}
where $V_0$ is the world volume. 
The argument (\ref{Z_Ca}) is 
different from our $Z$, Eq. (\ref{Z_C}) in the absence of the second term. 
This is due to the difference in the order of the two procedures:
quantization and reduction of the degrees of freedom.
In our case, the cosmological isotropic symmetry is 
taken after quantization, 
while it is taken before quantization in Ref. \cite{Horiguchi94}.
The property of quantum fluctuations depends on 
how to construct minisuperspaces.
Next we refer to the special case $m=Q \neq 0$.
The asymptotic line element is 
\begin{equation}
    ds_{dBB}^2 \to 
    -(1+\frac{m}{x})^{-2}dt^2
    +a(t)^2(1+\frac{m}{x})^2(d\rho^2+\rho^2d\Omega^2),
\end{equation}
where $x=a(t)\rho$.
This is called  an extreme black hole.
In classical theory, extension of the Majumdar-Papetrou geometry 
to the cosmological black hole 
was discussed by Ref. \cite{Kastor93}.  

%%%%%%%%%%%%%%%%%%%%%%%%%%%% case D  %%%%%%%%%%%%%%%%%%%%%%%%%%%%%%
\vspace{3mm}
\noindent
Case D: $\bar{\chi}(R)\equiv\frac{1}{2}(F(R)+\sqrt{F(R)^2+4R^2})$
\vspace{3mm}

The ansatz yields 
\begin{equation}
\Lambda^2 = {1 \over 2}(-F(R)+\sqrt{F(R)^2+4R^2})
\end{equation}
and $R' = R$ from Eqs. (\ref{solution Lambda}) and (\ref{dashR}), 
respectively.
Here $c_0=1$ has been chosen.
Thus the initial $R$ takes a simple form
\begin{equation}
R=\exp({r+\phi(t))},
\end{equation}
where $\phi(t)$ is an arbitrary function of $t$ as an integration constant 
with respect to $r$.

If we take the lapse function $N$ as
\begin{equation}
N^2={\lambda \over 6}\bigl( F(R)+\sqrt{F(R)^2+4R^2}~\bigr) n(Z)^2,
\end{equation}
Eq. (\ref{dBB R}) is reduced to
$\dot \phi =\sqrt{{\lambda \over 3}}$.
Therefore the form of $R$ is determined as 
\begin{equation}
R=a_0 \exp(\sqrt{{\lambda \over 3}}t+r)\equiv a(t)\rho(r) ,
\end{equation}
where $a_0$ is a constant.
Resultantly the line element $ds_{dBB}$ is given by  
\begin{eqnarray}
ds_{dBB}^2 
&=&-\frac{\lambda}{6}(\sqrt{F(R)^2+4R^2}+F(R)) n(Z_{dBB})^2dt^2 \nonumber \\
&+&a^2(t)\bigl( \frac{2 \, d\rho^2}{{\sqrt{F(R)^2+4R^2}+F(R)}}
        + \rho^2 d\Omega^2\bigl).
\label{ds_E}
\end{eqnarray}
The classical limit of Eq. (\ref{ds_E}) shows a cosmological black hole 
geometry in the standard form,
which contrasts with the isotropic form in the case C.

In the asymptotic region where $n_{dBB}\to 1$, 
or the classical limit $\hbar\to 0$,  
Eqs. (\ref{ds_B}), (\ref{ds_C}), and (\ref{ds_E}) show  
the same classical Reissner-Nordstr\"{o}m-de Sitter space-time.
They are different from each other merely in the coordinate system with 
respect to which they are described and the region which they cover in 
the fully extended space-time.
In Eq. (\ref{ds_B}) of the case B, the coordinate $t$ plays the role of 
the Schwarzschild radial coordinate.
The associated condition $F(R) \le 0$ shows that 
the corresponding classical geometry covers only the dynamical region, or 
one patch bounded by the horizons.
On the other hand, the asymptotic form of Eq. (\ref{ds_C}) 
is described with respect to the cosmological isotropic 
coordinate and covers the region from the inner of the event horizon of the
black hole to the outer of the event horizon of the de Sitter universe.
The transformation between the cosmological isotropic coordinate 
and the static one is given by 
\begin{equation}
a\Omega \rho =\bar{r},~ 
t=\bar{t}-\sqrt{{\lambda \over 3}}
    \int  {1 \over F(\bar{r})}
      {\bar{r}^2 \over \sqrt{\bar{r}^2-2m\bar{r}+Q^2}}
     d\bar{r}, 
\label{B-C trans}
\end{equation}
where $\bar{t}$ and $\bar{r}$ are the time and radial coordinates
 in the static Reissner-Nordstr\"{o}m-de Sitter space-time.
We find that the coordinate $t$ plays the role of an advanced null 
coordinate.
Alternatively, the coordinate transformation 
\begin{equation}
a\rho =\bar{r},~ 
t=
\bar{t}-\int  {1 \over F(\bar{r})}
      {\sqrt{F^2(\bar{r})^2+\bar{r}^2}-F(\bar{r}) \over \bar{r}} d\bar{r}
\label{B-E trans}
\end{equation}
relates the asymptotic form of Eq. (\ref{ds_E}) with that of 
Eq. (\ref{ds_B}) to extend the dynamical region to the static region.

Returning our attention to quantum theory, 
we find that the rigid space-time picture is not covariant under general 
coordinate transformations.
For example, under the coordinate transformation (\ref{B-E trans}), 
Eq. (\ref{ds_B}) is not equivalent to Eq. (\ref{ds_E}) 
due to the presence of the quantum gravity factor $n(Z_{dBB})$.
In other word, the quantum effect appears differently 
depending on the choice of coordinate systems.
It may appear strange.
Indeed, a quantum state on the superspace is constructed to be invariant.
However, the deterministic picture labels quantum fluctuations on the 
superspace by introducing the new time coordinate.
In this way, the dynamics is essential to quantum mechanics.

%Here we propose that the difference in the coordinate system and the 
%the region occurs due to the 
%initial configuration $(R(r), \Lambda(r))$ 
%rather than $N$ and $N^r$ chosen when we integrate the quantum 
%equations of motion.
%As circumstantial evidence, we can present 
%the other initial configuration 
%with the nonzero constant curvature.

By giving the initial and coordinate conditions concretely,
in this section, we obtained the various space-times 
which are related with each other by the coordinate transformations.
For other conditions, we can also perform analytic evaluations.
In Appendix, some calculations are given.
As a nontrivial case, 
we here present a space-time with the positive constant curvature.
For $m=Q=0$, the configuration variables $(R, \Lambda)=(a(t)r, 
a(t)/\sqrt{1-Kr^2})$ satisfies Eq. (\ref{rational2})  
when we set the shift function as $N^r=0$.
Here $K(>0)$ is a sign of the constant curvature.
The variable $a(t)$ is determined by Eq. (\ref{dBB R}) as 
\begin{equation}
\dot a={N \over n(Z)}\sqrt{\frac{\lambda}{3}a^2-K}.
\end{equation}
When we fix the lapse function as $N=n(Z)$, the solution is 
\begin{equation}
a(t)=\sqrt{\frac{3K}{\lambda}}
     \cosh{\sqrt{\frac{\lambda}{3}}t}.
\label{a_F}
\end{equation} 
In the classical limit $\hbar\to 0$, 
this solution corresponds to 
the closed de Sitter universe whose scale factor is Eq. (\ref{a_F}).
Now the corresponding classical geometry cannot be related with 
the cases A-D by any coordinate transformation.

%%%%%%%%%%%%%%%%%%%%%%%%%%%%%%%%%%%%%%%%%%%%%%%%%%%%%%%%%%%%%%%%%%
%
\section{Summary}
%
%%%%%%%%%%%%%%%%%%%%%%%%%%%%%%%%%%%%%%%%%%%%%%%%%%%%%%%%%%%%%%%%%%
We have studied canonical quantization of the Einstein-Maxwell 
theory with a cosmological term in spherically symmetric space-time
from the viewpoint of the de Broglie-Bohm interpretation. 
First we constructed the canonical formalism of the spherically 
symmetric geometry and quantized it.
To resolve the operator ordering ambiguity, we followed 
the procedure proposed by our previous work\cite{Kenmoku99}.
We obtained an analytic wave function, which is the simultaneous 
eigenstate of the mass and electric charge operators.
It is noted that, at this stage, 
any coordinate fixing had not been taken and the lapse and 
shift functions $N$ and $N^r$ were still arbitrary. 
Next the dBB interpretation of quantum mechanics 
was applied to the wave function, 
as if the Wheeler-DeWitt equation 
were the Schr\"{o}dinger equation with zero energy.
The equations of motion for the quantized space-time geometry were given as 
deterministic partial differential equations. 
Giving the ansatz to specify an initial geometry
 and choosing a coordinate system, 
we integrated the equations analytically.
In the asymptotic limit, the obtained rigid metrics correspond to 
the classical solutions of the Einstein-Maxwell theory, 
or the various representations of the 
Reissner-Norstr\"{o}m-de Sitter space-time and the closed de Sitter 
universe.

As discussed in Sec. 4 in detail, 
comparison between the rigid geometries as the dBB trajectories shows that 
the dBB trajectory picture is not covariant under 
coordinate transformations.
When we translate a quantum state from the wave function picture 
to the dBB trajectory representation,
we need additionally the coordinate system 
which describes the canonical evolution. 
If once we give a coordinate condition and obtain the rigid space-time 
picture,  
the metric of the space-time cannot be transformed to the others induced 
by the different choice of coordinate systems under any coordinate 
transformation.
In other words, 
realization of the quantum world breaks the covariance under 
general coordinate transformations, 
while  a quantum state as bundle of possible rigid geometries has 
invariance.
It may remind us the effective action at the spontaneous symmetry breaking.

In our analysis, a coordinate system is chosen a priori.
However, if it has the important meaning on the quantum gravity, 
it must be treated as physical process fundamentally.
In order to make our analysis more plausible, we have to throw probes 
into quantum-mechanically fluctuating space-times 
rather than choose a coordinate condition as a special one.
The probe will play the role of the coordinate system 
in the sense that it set up how to measure a space-time structure.
The dBB trajectory picture 
constructed by the information from the probe will 
teach us what quantum gravitational state is realized.
We may think the probe and the observable information as 
a quantum field and the Hawking radiation, for example.
The analysis on this line is needed for the conceptual problem.

There is another interest in our work from the angle of model analysis.
Using the model discussed here, we can construct the quantum mechanics of 
extreme black holes\cite{extreme}, especially of 
cosmological extreme black holes \cite{Kastor93}, 
and analyze the gravitational fluctuation near the event horizon 
at $N^r\ne 0$ gauge analogous to the Vaidya metric, 
which gives us a different viewpoint of the Hawking radiation
\cite{Hawking75}.

\appendix
\setcounter{section}{1}
\begin{center}
{\bf APPENDIX}
\end{center}

Following the analysis in Sec. 4, we present other calculations of 
the dBB trajectory.
%%%%%%%%%%%%%%%%%%%%%%%%%%%%%%%%%%%%%%%%%%%%%%%%%%%%%%%%%%%%%%%%%%
%
\subsection{Ansatz: $\bar{\chi}(R)\equiv -\frac{\lambda}{3}R^2$}
%
%%%%%%%%%%%%%%%%%%%%%%%%%%%%%%%%%%%%%%%%%%%%%%%%%%%%%%%%%%%%%%%%%%
We give the ansatz $\bar{\chi}(R)\equiv -\frac{\lambda}{3}R^2$ and 
treat only the region $\bar \chi (R)-F(R) <0$.
Taking $c_0(r)=1$ in Eq. (\ref{fbar}),
we find
\begin{equation}
\Lambda^2 = -(1-\frac{2m}{R}+ \frac{Q^2}{R^2}).
\end{equation}
Here we allow $\Lambda$ to become purely imaginary only if 
the reality of $Z$ is reserved.
From Eq. (\ref{solution R}), we obtain 
\begin{equation}
R=y(1+\frac{m}{y}+\frac{m^2-Q^2}{4y^2}) \ ,
\end{equation}
where $y\equiv\exp{(\sqrt{{\lambda \over 3}}r+\phi(t))}$, 
and $\phi(t)$ is an arbitrary function of t as an integration constant.
If we adopt a coordinate condition
\begin{equation}
N^2=-y^2(1+{m \over y}+{m^2-Q^2 \over 4y^2})^2 n(Z)^2,
\end{equation}
where the evolution of the variables is traced on the 
purely imaginary time $t$, 
Eq. (\ref{dBB R}) is reduced to $\dot{\phi}=1$, and thus 
$y =a_0\exp{(\sqrt{{\lambda \over 3}}r+t)}$, where $a_0$ is a constant.
As a result, the line element $ds_{dBB}$ is
\begin{equation}
ds_{dBB}^2=-\left (\frac{1-\frac{m^2-Q^2}{4y^2}}
    { 1+\frac{m}{y}+\frac{m^2-Q^2}{4y^2}}\right)^2 dr^2
     +y^2\left(1+\frac{m}{y}+\frac{m^2-Q^2}{4y^2}\right)^2
       (n(Z_{dBB})^2 dt^2+d\Omega^2).
\label{ds_D}
\end{equation}

%%%%%%%%%%%%%%  space-time dual transformation  %%%%%%%%%%%%%%
%
%By imposing the gauge condition for 
%$N^r$ and  coordinate conditions 
%for $R'$ through the choice of $\chi$ and $\dot R$, 
%the quantum trajectory is in general fixed and approaches 
%to various classical solutions in the asymptotic region. 
%These classical solutions are related by general coordinate 
%transformations.  
For $N^r=0$, we define a space-time dual transformation as 
\begin{eqnarray}
    r \rightarrow t \ , \ \  t \rightarrow r \ , \ \ 
    \Lambda \rightarrow iN \ \ {\rm and} \ \  N \rightarrow -i\Lambda \ .  
\end{eqnarray} 
Under this transformation, 
the action (\ref{I-ADM2}) is invariant and the 
classical solutions (asymptotic forms) derived in Sec. 4 
are related with each other: 
\begin{eqnarray}
    {\rm case~A} \leftrightarrow {\rm case~B} \ \ \ {\rm and} \ \ \ 
    {\rm case~C} \leftrightarrow {\rm Appendix} \ .
\end{eqnarray}
However the dBB trajectory representation is not 
covariant under the space-time dual 
transformation.

%%%%%%%%%%%%%%%%%%%%%%%%%%%%%%%%%%%%%%%%%%%%%%%%%%%%%%%%%%%%%%%%%%
%
\subsection{Ansatz: $\bar{\chi}(R)\equiv 0~~(N^r\ne 0)$}
%
%%%%%%%%%%%%%%%%%%%%%%%%%%%%%%%%%%%%%%%%%%%%%%%%%%%%%%%%%%%%%%%%%%
In Sec. 4, we use the shift function $N^r=0$ as a coordinate condition.
Here we present an example of the nonzero shift function.
%\begin{equation}
%N^r=N/\Lambda.
%\end{equation}
From the ansatz $\bar{\chi}(R)\equiv 0$, we find $R'=0$ and that  
Eq. (\ref{Zint}) is reduced to
\begin{equation}
Z=\int dr \Lambda R\sqrt{-F(R)}.
\label{Z_App}
\end{equation}
By estimating the  derivative of $Z$ with respect to $R$, we have 
\begin{equation}
{\delta \Theta \over \delta R}
=\Lambda \bigl(\sqrt{-F(R)}+R\frac{d}{dR}\sqrt{-F(R)}\bigr)
                                                   {d\Theta \over dZ}.
\end{equation}
The equations of motion for the dBB trajectory are reduced to
\begin{equation}
\dot R={N \over n(Z)}\sqrt{-F(R)}{\rm ~~~~~and~~~~~}
\dot \Lambda-(N^r\Lambda)'={N \over n(Z)}\Lambda {d\over dR}\sqrt{-F(R)}.
\end{equation}
If we set $N = n(Z)/\sqrt{-F(R)}$ and $(N^r)'=0$, then we find 
$R=t$ and 
\begin{equation}
\Lambda=G(r+\int N^r dt)\sqrt{-F(t)},
\end{equation}
where G is a arbitrary function.
Thus the line element $ds_{dBB}$ is
%\begin{equation}
%ds^2_{dBB} =-F(t)dr^2+ 2\frac{n(Z_{dBB})}{F(t)}dtdr +t^2d\Omega^2.
%\label{ds_App}
%\end{equation}
\begin{equation}
ds^2_{dBB}=\frac{n(Z_{dBB})}{F(t)}dt^2
-F(t)G(r+\int N^r dt)^2(dr+N^rdt)^2+t^2d\Omega^2,
\label{ds_App}
\end{equation}
where
\begin{equation}
Z_{dBB}=t\sqrt{-F(t)}\int dr G(r+\int N^r dt).
\end{equation}
Equation (\ref{ds_App}) can be related with Eq. (\ref{ds_B}) by the 
coordinate transformation $\bar{t}=t$ and $\bar{r}=\int G(\alpha)d\alpha$,
where $d\alpha=dr+N^rdt$, and $\bar{t}$ and $\bar{r}$ denote the 
coordinates in the static Reissner-Norstr\"{o}m-de Sitter metric.
%This asymptotic metric corresponds to the Vaidya metric 
%when we exchange $t$ and $r$.

\clearpage
%%%%%%%%%%%%%%%%%%%%%%%%%%%%%%%%%%%%%%%%%%%%%%%%%%%%%%%%%%%%%%%%%%%%%%%%%%

\clearpage
\pagebreak
\end{document}